\documentclass[fleqn,usenatbib]{mnras}

\usepackage[utf8]{inputenc}
\usepackage{amsmath}
\usepackage{graphicx}
\usepackage{epstopdf}
 \usepackage{auto-pst-pdf} 
\usepackage{threeparttable}
\usepackage{graphicx, color}
\usepackage{bm}
\usepackage{soul}
\usepackage{amssymb}
\usepackage{hyperref}

\newcommand{\eqb}{\begin{eqnarray}}
\newcommand{\eqe}{\end{eqnarray}}
\newcommand{\ergs}{erg~s$^{-1}$}

\newcommand{\Le}{L_{\rm e}}
\newcommand{\Lbr}{L_{\rm br}}
\newcommand{\tbmax}{t_{\rm b, max}}
\newcommand{\temin}{t_{\rm e, min}}
\newcommand{\tb}{t_{\rm b}}
\newcommand{\te}{t_{\rm e}}
\newcommand{\tg}{t_{\gamma}}
\newcommand{\pb}{p_{\rm b}}
\newcommand{\pe}{p_{\rm e}}
\newcommand{\pg}{p_{\gamma}}

\title[Central engines of collapsar GRBs]
{Collapsar Gamma-Ray Bursts: how the luminosity function dictates the duration distribution}
\author[Petropoulou, Barniol Duran, Giannios]{Maria Petropoulou, Rodolfo Barniol Duran \& Dimitrios Giannios\\
Department of Physics, Purdue University, 525 Northwestern Avenue, West Lafayette, IN, 47907, USA \\
}

\begin{document}

\date{Received.../Accepted...}

\pagerange{\pageref{firstpage}--\pageref{lastpage}} \pubyear{2017}

\maketitle

\label{firstpage}

\begin{abstract}
Jets in long-duration $\gamma$-ray bursts (GRBs) have to drill through the collapsing 
star in order to break out of it and produce the $\gamma$-ray signal while the central engine
is still active. If the breakout time is shorter for more powerful engines, then the jet-collapsar interaction acts
as a filter of less luminous jets. We show that the observed broken power-law GRB luminosity function is a natural outcome of this 
process. For a theoretically motivated breakout time that scales with jet luminosity as $L^{-\chi}$ with $\chi\sim 1/3-1/2$, we show that the shape of the $\gamma$-ray duration distribution can be uniquely determined by the GRB luminosity function and matches the observed one. This analysis has also interesting implications about the supernova-central engine connection. We show that not only successful jets can deposit sufficient energy in the stellar envelope to power the GRB-associated supernovae, but also failed jets may operate in all Type Ib/c supernovae. 
\end{abstract}


\section{Introduction}
The central engine of $\gamma$-ray bursts (GRBs, see, \citealp[e.g.][]{kumarandzhang2015} for a review) is 
hidden to direct observation. However, its workings may be imprinted in observational signals 
of this phenomenon. \cite{bromberg2012,bromberg2013} proposed that the prompt $\gamma$-ray duration distribution 
of collapsar GRBs exhibits a plateau, indicating the fact that GRB jets launched at the core of a 
collapsing star must drill their way out of the star and break out from its surface before producing the observed 
$\gamma$-rays (for the collapsar model, see,
\citealp[e.g.][]{paczynski1998, macfadyen1999}). This claim can be 
extended to understand low-luminosity GRBs as jets that barely failed to
break out \citep{bromberg2011}. Recently, these arguments 
have been used to suggest that failed jets may operate in all Type Ib/c supernovae (SNe) \citep{sobacchi2017}.

Previous works have focused on the $\gamma$-ray duration distribution and assumed a single breakout time 
for all collapsar GRBs \citep[e.g.,][]{sobacchi2017}. However, differences in the properties of long-duration GRB engines
should yield different breakout times. In particular, one expects that more powerful jets will propagate more easily 
through the star and will break out from it much quicker than weaker jets \citep[e.g.][]{zhang2003, morsony2007, mizuta2009, lazzatietal2012}. 
Both analytical estimates \citep{bromberg2011, bromberg2011b} and
numerical simulations \citep[see][and references
therein]{lazzatietal2012} suggest that the jet's breakout time
depends upon the engine's isotropic luminosity as $\Le^{-\chi}$; the
power-law index lies between 1/3 and 1/2 depending on properties of
the stellar envelope (e.g., density profile, radius and mass) and/or
the properties of the jet (e.g., jet composition). 

In this paper, we consider the luminosity dependence of the breakout time in a scenario where the jet needs to 
drill through the collapsing star before it can power a GRB. An extended distribution in engine luminosities is motivated by the GRB luminosity function itself, which extends several orders of magnitude in isotropic luminosity \citep[e.g.,][]{wanderman_piran2010}. We show that a broken power-law GRB luminosity function is an expected outcome of the jet-envelope interaction for central engines having a single power-law luminosity distribution.  After matching the parameters of the model-predicted GRB luminosity function to the observed one, we derive a {\sl mono-parametric} $\gamma$-ray duration distribution. The maximum breakout time (i.e., its single tunable parameter) can be inferred by comparison to the observed GRB collapsar duration distribution. This framework is quite powerful as it connects observable quantities with 
the properties of the GRB central engine and of the collapsing star. It also makes a tantalizing connection between GRBs and jet-driven 
core-collapse supernovae (e.g., \citealp{soderberg2010, lazzatietal2012, marguttietal2014, sobacchi2017, piranetal2017, bear2017}, for a review, see, e.g., \citealp{soker2016} and references therein).

\section{Model setup}
We consider a scenario similar to that presented in \cite{bromberg2012}. A long-duration GRB central engine that launches a jet\footnote{Hence, we use the terms central engine and (injected) jet interchangeably.} for 
a duration $\te$ must be active for a duration longer than the time $\tb$ it takes the jet to break out of the stellar 
envelope in order to produce a $\gamma$-ray signal. The duration of the prompt $\gamma$-ray emission is thus 
$\tg = \te - \tb$. For $\te < \tb$ the jet is unable to break out from the star and ``fails'', but still injects its energy to 
the stellar envelope, possibly powering the supernova explosion (see section 4). Contrary to previous studies 
\citep{bromberg2012, sobacchi2017}, we consider a distribution of engine luminosities which, in turn, translates 
into (i) a distribution of observed GRB luminosities and (ii) a distribution of breakout times. 
In this section we describe our assumptions regarding the distribution of engine luminosities and their duration. 

We consider a normalized power-law distribution of isotropic engine luminosities $\Le$ as
 \eqb
 \label{eq:pL}
 p(\Le)\equiv \frac{{\rm d}N}{{\rm d}\Le}=\frac{a-1}{L_{\rm e, \min}}\left( \frac{\Le}{L_{\rm e, \min}}\right)^{-a}, \ a> 1,
 \eqe 
 where $L_{\rm e,min}$ is the engine's minimum isotropic luminosity. Whether or not the 
 jet breaks out from the star depends on the breakout time $\tb$, which is related to $\Le$ as:
 \eqb
 \label{eq:tb}
 \tb=t_0 \left(\frac{\Le}{L_{\rm e,0}} \right)^{-\chi}.
 \eqe 
 where $L_{\rm e,0}=10^{51}$~\ergs. Henceforth, we consider a narrow
 range of possible power-law indices that are motivated by relativistic hydrodynamic simulations of jet propagation in collapsars,
 namely $1/3 \lesssim \chi \lesssim 1/2$ \citep[e.g.][]{bromberg2011, lazzatietal2012, nakar2015}. The normalization $t_0$ encodes information about the jet's collimation and the properties of the stellar envelope. In our analysis, we will treat $t_0$ as a free parameter and assume it is the same among all GRB collapsars.
  
 The distribution of breakout times, $\pb(\tb)$, can be determined by the condition $\pb(\tb) {\rm d}\tb = p(\Le) {\rm d} \Le$ using equations (\ref{eq:pL}) and (\ref{eq:tb}). The resulting distribution can be written as:
 \eqb
 \label{eq:pb}
  \pb(\tb) & = &  A_{\rm b} \left( \frac{\tb}{t_0} \right)^{s}H[\tbmax-\tb] 
 \eqe 
 where $H[x]$ is the Heaviside step function,   $s=(a-1-\chi)/\chi$, and 
 \eqb 
 A_{\rm b} & = & \frac{a-1}{\chi t_0}\left(\frac{L_{\rm e,0}}{L_{\rm e,min}}\right)^{-a+1}.
 \eqe
 The maximum breakout time $\tbmax$ corresponds to an engine with the minimum luminosity and is given by:
 \eqb 
 \tbmax & = & t_0\left(\frac{L_{\rm e,min}}{L_{\rm e,0}}\right)^{-\chi}. 
 \label{eq:tbmax}
 \eqe 
 The distribution of engine durations $\te$ is assumed to be unrelated to the distribution of engine luminosities $\Le$ and to follow a power law, similar to previous studies \citep[e.g.,][]{bromberg2012, sobacchi2017}:
 \eqb
 \label{eq:pe}
 \pe(\te) = A_{\rm e} \left( \frac{\te}{\temin}\right)^{-\beta}H[\te - \temin], \ \beta > 1,
 \eqe 
 where  $\te =\tg+\tb$ and  $\temin$ is the minimum duration of the central engine. The normalization is 
 \eqb
 A_{\rm e}=\frac{\beta-1}{\temin}
 \eqe 
 and ensures that $\int_0^{\infty}{\rm d}\te \pe(\te)=1$.  

\section{Results}
In this section we connect the engine distributions described previously with the observed distributions of luminosities and  durations
of long (collapsar) GRBs.
 
Of all possible central engines, only those with $\te > \tb$ can lead to ``successful''  GRBs, that is, 
jets that can break out from the star and produce a $\gamma$-ray signal while the engine is still active. 
The fraction of such successful jets is given by:
\eqb
f(\tb) = \int_{\tb}^{\infty} {\rm d}\te \pe(\te)
\eqe 
and can be written as:  
\eqb
\label{eq:fout}
f(\tb; \Le) = \left(\frac{\tb}{\temin} \right)^{-\beta+1}H[\tb - \temin] + H[\temin - \tb],
\eqe 
where the dependence upon $\Le$ comes through $\tb$. The last term in
the right hand-side of the equation shows that 
all engines with breakout times shorter than the minimum duration $\temin$ will be successful in producing GRBs.

\subsection{Intrinsic GRB luminosity function}
We argue that the (normalized) isotropic GRB luminosity function is 
the product of the fraction $f$ of successful jets and the engine luminosity function $p(\Le)$: 
\eqb 
\frac{{\rm d}N_{\rm GRB}}{{\rm d}L} = \frac{f(\tb;\Le)}{\eta}\frac{{\rm d}N}{{\rm d}\Le},
\eqe 
where we assumed that the  isotropic GRB (radiated) luminosity $L$ is a constant fraction of the 
isotropic engine luminosity (i.e., $L=\eta \Le$), as motivated by the narrow distribution of the $\gamma$-ray efficiency of GRBs \citep[e.g.][]{fan_2006, beniamini_2016}. Depending on whether $L$ denotes the peak or average burst luminosity, the numerical value of $\eta$ will differ by a factor of $\sim 3$. 

Using equations (\ref{eq:pL}), (\ref{eq:tb}), and (\ref{eq:fout}) we find:
\eqb
\label{eq:pGRB}
\frac{{\rm d}N_{\rm GRB}}{{\rm d}L} =N_0\left(\frac{L}{L_{\min}}\right)^{-a}
\left\{ \begin{array}{l}
\left(\frac{L}{\Lbr}\right)^{-w}, L \le \Lbr \\ \\
1, \ L > \Lbr, 
\end{array}
\right.
\eqe 
where $N_0\equiv (a-1)/L_{\min}$,  $w=\chi(-\beta+1)<0$ and the break luminosity is defined as:
\eqb 
\label{eq:Lbr}
\Lbr \equiv L_0 \left(\frac{\temin}{t_0} \right)^{-1/\chi}.
\eqe 
Therefore, a power-law distribution of the central engine luminosity
results in a broken power-law distribution of the GRB luminosity
function. The break at $L_{\rm br}$ is due to the fact that an
increasing fraction of  engines with lower luminosities are not able
to produce successful GRBs. In contrast, the GRB luminosity function reflects the luminosity function of the central engine for $L > \Lbr$.

\begin{figure}
 \centering 
\includegraphics[trim={0cm 0cm 0cm 0.8cm},clip,width=0.49\textwidth]{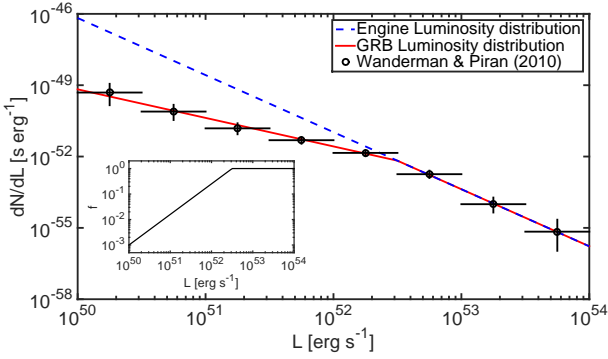}  
\caption{The distribution of engine luminosities, ${\rm d}N/{\rm d}L = (1/\eta){\rm d}N/{\rm d}\Le$, assumed to be a power-law (blue dashed line), experiences a break as we consider only jets that can break out of the star to produce a distribution of successful GRBs, ${\rm d}N_{\rm GRB}/{\rm d}L$ (red solid line). We match this distribution to that found by \protect\cite{wanderman_piran2010}. A similar matching can be done for other luminosity functions. The inset shows the fraction of successful jets as a function of luminosity, see equation (\ref{eq:pGRB}).}  
\label{fig:fig1}
\end{figure}

Observationally, the GRB luminosity function is, indeed, best described by a broken power law \citep[e.g.][]{wanderman_piran2010, salvaterra2012, pescalli2016}:
\eqb
\label{eq:pgrb_intrinsic}
\phi(L) \equiv \frac{{\rm d}N_{\rm GRB}}{{\rm d}\log L} = \left \{ 
\begin{array}{ll}
\left(\frac{L}{L_*}\right)^{-\alpha_L} & L \le L_*  \\
\left(\frac{L}{L_*}\right)^{-\beta_L} & L > L_*.
\end{array}
\right.
\eqe
Comparison of the GRB luminosity function with the predicted one, equation (\ref{eq:pGRB}), allows us to uniquely determine the critical parameters of our model that describe the engine luminosity distribution and its duration:
\eqb 
\Lbr & = &  L_* \label{eq:Lstar} \\ 
a & = & \beta_L +1 \label{eq:a} \\ 
\beta & = & \frac{\beta_L-\alpha_L}{\chi}+1 \label{eq:beta} \\
\temin & = & t_0 \left(\frac{10 L_*}{\eta_{-1} L_{\rm e,0}} \right)^{-\chi} \label{eq:temin},
\label{eq:param}
\eqe 
where we use the standard $Q_{\rm x} = 10^{-x} Q$ notation. We present a graphical view of our comparison in Fig.~\ref{fig:fig1} and  summarize our results for the two indicative $\chi$ values in Table~\ref{tab:tab1}.  Interestingly, the minimum engine activity timescale cannot be arbitrarily short for a given $t_0$.

\begin{table}
\centering
\caption{Model parameters for jet propagation through the stellar envelope as determined by the GRB 
luminosity function and the collapsar GRB duration distribution. Results for two theoretically motivated $\chi$ 
values are shown. The power-law indices are defined in equations (\ref{eq:pL}) and (\ref{eq:pe})
$\tbmax$ is obtained from the observed $\gamma$-ray duration 
distribution, see Fig. \ref{fig:fig2}. The parameter values of the GRB luminosity function are fixed to the best-fit values of \citet{wanderman_piran2010}. 
}
\begin{tabular}{ccccc}
\hline 
 $\chi$   &  $a$ & $\beta$ & $\temin$ & $\tbmax$ \\
\hline
1/3& 2.4 & 4.6 & $t_0/7$ & 69~s \\
1/2& 2.4 & 3.5 & $t_0/18$& 47~s \\      
\hline
\end{tabular}
 \label{tab:tab1}
\end{table}
\subsection{Distribution of $\gamma$-ray durations}

The distribution of rest-frame $\gamma$-ray durations of GRBs $\tg = \te - \tb$ can be calculated using the procedure outlined 
in \cite{sobacchi2017}. However, we relax their assumption of a fixed
$\tb$ value by considering a distribution of breakout times as determined by
the jet luminosity. In this case, $\pg$ is calculated as:
\eqb 
\label{eq:pg-def}
\pg(\tg) = \int_0^{\infty} {\rm d}\tb \pb(\tb) \frac{\pe(\tg+\tb)}{f(\tb; L_e)}.
\eqe
The above equation can be recast in the form:
\eqb
\label{eq:pg}
\pg(\tg)& = & A_{\rm b}A_{\rm e}t_0^{-s}\left(\frac{\tg}{\temin}\right)^{-\beta}\left[I_1(\tg) + I_2(\tg) \right] \\ 
I_1(\tg)& = & \int_{\temin-\tg}^{\temin} {\rm d}\tb \tb^s \left(1+\frac{\tb}{\tg}\right)^{-\beta} \\
I_2(\tg)& = & \temin^{1-\beta}\int_{\temin}^{\tbmax} {\rm d}\tb  \tb^{s+\beta-1} \left(1+\frac{\tb}{\tg}\right)^{-\beta}
\eqe
 where $\tbmax$ corresponds to the breakout time of the minimum isotropic engine luminosity $L_{\rm e, min}$, 
 see equation (\ref{eq:tbmax}).  The distribution of $\gamma$-ray durations can be analytically derived in two limiting regimes \citep[see also][]{bromberg2012}:
\begin{enumerate}
 \item $\tg \ll \tb$, where $\te \approx \tb$.  Using also $\tbmax \gg \temin$, equation (\ref{eq:pg}) can be cast in the form: 
 \eqb
 \label{eq:pgshort}
 \pg(\tg) \simeq \frac{\beta_L(\beta_L-\alpha_L)}{\chi(\beta_L-\chi)}\tbmax^{-1},
 \eqe 
 which is independent of $\tg$.  For short enough
 $\gamma$-ray durations, $\pg$ is set by the maximum breakout
 timescale and the power-law indices of the GRB luminosity function. 
 \item $\tg \gg \tb$, where $\te \approx \tg$. Equation (\ref{eq:pg-def}) then results in:
  \eqb
  \label{eq:pglong}
  \pg(\tg) \simeq \frac{\beta_L(\beta_L-\alpha_L)}{\chi (2\beta_L-\alpha_L) \tbmax}\left( \frac{\tg}{\tbmax}\right)^{-1-\frac{\beta_L-\alpha_L}{\chi}}.
  \eqe 
In this regime, the distribution of durations reflects the
distribution of engine timescales ($\pg \propto \tg^{-\beta}$), in agreement with previous works
\citep{bromberg2012, sobacchi2017}. However, in our study, the
power-law slope of the engine time distribution $\beta$ is {\it not} a free
parameter to be determined by comparison to the GRB duration distribution. It is instead
uniquely determined by the power-law slopes of
the  GRB luminosity function (i.e. $\alpha_L, \beta_L$)   and $\chi$
and can be used as a test of the model against the observed duration distribution of GRBs.
\end{enumerate}

For a specific jet propagation theory ($\chi$ value) and GRB luminosity function,  
the only free parameter which enters the calculation of $\pg(\tg)$ is the maximum breakout time. This affects both the ``plateau'' and the turnover  of the distribution $\pg$, see equations (\ref{eq:pgshort}) and (\ref{eq:pglong}). Thus, $\tbmax$ can be determined by fitting the model, equation (\ref{eq:pg}), to the duration distribution of collapsar GRBs.

To derive the observed GRB duration distribution, we used the $T_{90}$\footnote{This is a measure of the GRB duration and is defined as the time interval during which 90 per cent of the photon counts have been detected \citep{kouveliotou1993}.} value of 1066 GRBs detected by the {\it Swift} satellite during the period of 2004 -- 2017\footnote{\url{https://swift.gsfc.nasa.gov/archive/grb_table/}}. Redshift information is available only for $\sim 1/3$ of the sample, with a median redshift of  $z\simeq 1.7$. This was adopted as the typical redshift of the entire {\it Swift} sample in order to construct the histogram of rest-frame $\gamma$-ray durations as $t_{90}=T_{90}/(1+z)$.  

\begin{figure}
 \centering
\includegraphics[width=0.49\textwidth, trim=20 100 20 80]{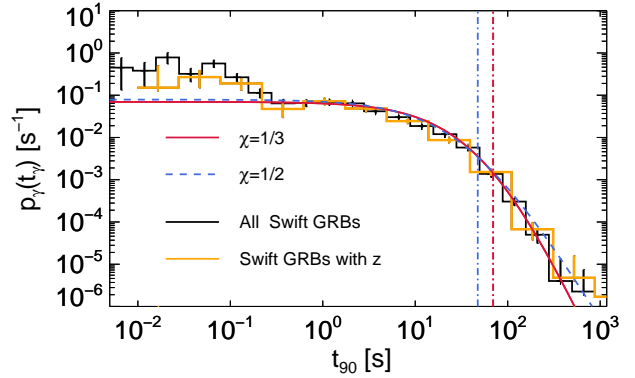} 
\caption{Histogram (normalized) of rest-frame $\gamma$-ray durations of all {\it Swift} GRBs (black line) 
assuming a typical redshift of $\sim 1.7$. 
The corresponding histogram of the 319 {\it Swift} GRBs with measured redshift is also shown (orange line).
Short GRBs ($t_{90} \lesssim 0.3$~s) do not have a collapsar origin and are therefore neglected in this study.
The model predictions for $\chi=1/3$ and $\chi=1/2$ are overplotted
for comparison. The model achieves a good quantitative  description of the data by adjusting its single parameter: the maximum
  breakout time $\tbmax$ (shown with dashed vertical lines). The results are obtained for the GRB luminosity function of \citet{wanderman_piran2010}.
} 
\label{fig:fig2}
\end{figure}

The normalized histogram of $t_{90}$ of all {\it Swift} GRBs is plotted in Fig.~\ref{fig:fig2} with a black line. The corresponding histogram of {\it Swift} GRBs with measured redshifts is also shown for comparison (orange line).  Overall, the two histograms are qualitatively similar with the main differences appearing at short durations. Regardless, GRBs with $t_{90} \lesssim 0.3$~s (i.e., short GRBs) do not have a collapsar origin and are neglected in this study. Overplotted as coloured curves  are our  model 
predictions with $\tbmax\simeq 69$~s for $\chi=1/3$ and $\tbmax \simeq 47$~s for $\chi=1/2$ (see also Table~\ref{tab:tab1}). Both provide 
very good quantitative descriptions to the collapsar GRB duration distribution. In fact, the plateau of the duration distribution is expected to be similar for models with the same product of $\chi$ and $\tbmax$, as long as $\beta_{\rm L} \gg \chi$. More specifically, $\pg \propto (\chi \tbmax)^{-1}$ at short enough durations, see equation (\ref{eq:pgshort}). From the ``fit'' to the duration distributions with $\chi=1/3$ and $\chi=1/2$ we find $\chi \tbmax \approx T_{\max}$ where $T_{\max}\simeq 23$~s.

A timescale of $\sim 60$~s -- beyond which the duration distribution turns into a steep power law -- is also found by \cite{sobacchi2017}. The authors associate this timescale with a {\it single} breakout time assumed for all 
GRBs, whereas in our framework it is clearly related to the maximum breakout time. In addition, 
\cite{sobacchi2017} fit their data in order to derive $\beta$, whereas in our model $\beta$ is {\it uniquely} determined 
by the GRB luminosity function and $\chi$ -- see equation (\ref{eq:pglong}). A fit to the data is not guaranteed in our model. For instance, the observed $\pg$ distribution 
cannot be explained by luminosity functions with $\beta_L-\alpha_L < 1$ for $\chi$ in the range $1/3-1/2$,  as motivated by theory
\citep{bromberg2011, lazzatietal2012}.

The maximum breakout time inferred from the observed GRB duration distribution can also be translated to a constraint between $t_0$ and a minimum luminosity for a given $\chi$ value. Using $\chi \tbmax\approx T_{\max}$ together with equation (\ref{eq:tbmax}), we find:
\eqb 
\label{eq:constraint-1}
T_{\rm max} \approx \chi \tbmax = \chi t_0 \left(\frac{L_{\min}}{\eta L_{\rm e,0}} \right)^{-\chi}.
\eqe

\section{Discussion}
The shape of the $\gamma$-ray duration distribution of collapsar  GRBs (i.e., plateau followed by a power-law) 
is generic and the result of: (i) the relation between the three
timescales, i.e. $\tg = \te - \tb$, which stems from the jet-stellar
envelope interaction, (ii) {\it at least} one power-law distribution
(either of $\tb$ or $\te$), and (iii) a characteristic 
timescale set by either $\tb$ or $\te$. The plateau in $\pg$ at short
durations is an even more general result of the jet-envelope
interaction model, as it only requires that $\pe$ is a smooth function of $\te$ \citep{bromberg2012}.

\subsection{Completeness of the GRB sample}

In our analysis we assumed that the {\it Swift} GRB sample is complete with respect to
the observed burst duration. In other words, for a fixed luminosity the detection of the burst does not depend on its duration. 
This is a simplifying assumption, for the minimum detectable flux of a GRB depends upon the exposure time $T$ as $T^{-1/2}$ \citep{baumgartner_2013, lien_2016}. For GRBs, the maximum exposure time is set by their duration. For simplicity, let us identify 
$T$ as the observed duration $T_{90}$. This implies that the flux limit for detection is higher for GRBs with shorter duration. Indeed, 
there is lack of luminous and short bursts (i.e., $L_{\rm iso} > 10^{52}$ erg s$^{-1}$ and $T_{90} < 2$~s) in the third 
{\it Swift}/BAT GRB catalog (see Fig.~25 in \citealp{lien_2016}). This can be the result of a genuine lack of GRBs with these properties 
or of unavailable redshift measurements. Regardless, these results suggest that the {\it Swift}/BAT sample is complete with respect to burst duration for long GRBs. Therefore, we do not expect our main conclusions to depend strongly on this assumption.

\subsection{Progenitors of collapsar GRBs}

For a distribution of breakout times we showed that the only free parameter that controls $\pg(\tg)$ 
(both the plateau and position of the break) is the maximum breakout time. This, in turn, depends upon 
$t_0$ and $L_{\min}$, see equations (\ref{eq:tbmax}) and (\ref{eq:constraint-1}). 
Here, $L_{\min}$ should not be interpreted as the intrinsic GRB minimum luminosity. In fact, 
the GRB sample becomes progressively  less complete for lower isotropic GRB luminosities \citep[e.g.,][]{wanderman_piran2010}. Thus, 
the duration distribution probes GRBs with isotropic luminosities down to an ``effective'' minimum luminosity $L_{\rm eff}$.
Substitution of $L_{\rm eff}\simeq 10^{51}$ erg s$^{-1}$ 
to equation (\ref{eq:constraint-1}) leads to a conservative constraint:

\eqb 
\label{eq:constraint}
\frac{t_0}{T_{\max}} \approx \frac{10^{\chi}}{\chi}\left(\frac{L_{\rm eff, 51}}{\eta_{-1}} \right)^{\chi},
\eqe 
which translates to $t_0\sim 150$~s for all $\chi$ values in the range $1/3-1/2$ and $T_{\max}\sim 23$~s. Values of $t_0 \gg 20$~s cannot be easily reconciled with the scenario of jet propagation through a compact progenitor, because of the weak dependence that $t_0$ has on stellar properties \citep[see][]{bromberg2011, bromberg2012}. Our analysis suggests the presence of an extended low-mass envelope surrounding the GRB progenitor, as concluded independently by \cite{sobacchi2017}. The weak dependence of the breakout time on $L_{\rm eff}$ makes this conclusion quite robust.

Assuming a common $t_0 \sim 150$~s for all GRB collapsars, we can further predict that 
the duration distribution will extend towards longer durations (i.e., larger $\tbmax$ and $T_{\max}$) as the 
sample becomes more complete at lower luminosities thanks to more sensitive future missions. Falsification of this prediction would provide indirect evidence for a distribution of $t_0$ among GRB progenitors, which lies beyond the scope of this paper.
\subsection{Energetics of central engines}
We next examine the energetics of central engines and their potential in powering a supernova explosion. All engines with 
$L_{\rm e} > L_{\rm e, *}\equiv \eta^{-1} L_*$ produce successful GRBs. The associated energy deposited to the 
stellar envelope by successful engines is  
\eqb 
\label{eq:Ees}
E_{\rm e, s} \approx f_{\rm b} \Le \tb(\Le)  \approx E_{\rm e,*} \left(\frac{\Le}{L_{\rm e, *}}\right)^{1-\chi},
\eqe
where $f_{\rm b} = 0.01 \ f_{\rm b, -2}$ is the jet beaming fraction and  
\eqb
E_{\rm e, *}\equiv f_{\rm b} L_{\rm e,*} \temin.
\eqe 
The minimum engine timescale $\temin$ is given by equation (\ref{eq:temin}). Failed jets can also inject a considerable amount of energy to the stellar envelope. For $L_{\rm e, min} \le L_{\rm e} \le  L_{\rm e,*}$ the engine activity timescale $\te$ lies between $\tbmax$ and $\temin$. The latter is still a good proxy for the average $\te$ in the range of $[\temin, \tbmax]$ because of  the steepness of $\pe(\te)$. The energy of a failed GRB engine is then given by 
\eqb 
\label{eq:Eef}
E_{\rm e,f} \approx E_{\rm e,*} \frac{\Le}{L_{\rm e,*}}.
\eqe 

\begin{figure}
 \centering 
\includegraphics[width=0.49\textwidth, trim=15 15 0 0]{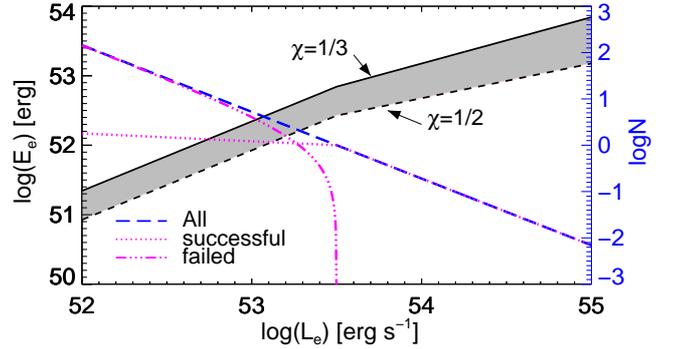}  
\caption{Energy deposited to the stellar envelope by successful and failed GRB engines for $\chi=1/3-1/2$ and $t_0=150$~s as obtained by our analysis (shaded region). Also shown are the number of successful (magenta dotted line) and failed (magenta dash-dotted line) engines as a function of engine luminosity. The total number of engines is overplotted (blue dashed line); a normalization of one engine with $L_{\rm e, *}= \eta_{-1}10^{53.5}$~\ergs was used.} 
\label{fig:fig3}
\end{figure}
Figure~\ref{fig:fig3} presents the beaming-corrected energy, $E_{\rm e}$, of successful and failed engines (see equations (\ref{eq:Ees}) and (\ref{eq:Eef}), respectively) as a function of $\Le$ for $\chi=1/3-1/2$ and $t_0=150$~s. For a typical GRB engine (i.e., $\Le = L_{\rm e,*}$), we find
\eqb 
\label{eq:Ees-num}
E_{\rm e,*} \approx (2-7) \times 10^{52} f_{\rm b,-2} \left(\frac{L_{*, 52.5}}{\eta_{-1}}\right)^{1-\chi} \, {\rm erg}
\eqe 
for $\chi = 1/3 - 1/2$.  This is comparable to the kinetic energy inferred in SNe associated to 
long GRBs \cite[e.g.,][]{hjorthandbloom2012} and provides further support that GRB-SNe are jet-driven \cite[e.g.,][]{hjorth2013}.  
Failed jets that inject $\sim 10^{51}$ erg (1 foe) to the stellar envelope -- an energy comparable to the 
binding energy of the stellar envelope and the typical kinetic energy of core-collapse SNe -- could facilitate the 
explosion mechanism in these SNe \citep{sobacchi2017, piranetal2017}, but they are not expected to drive mildly relativistic SNe ejecta \citep[see e.g.][]{soderberg2010}.  The isotropic engine luminosity in this case is $L_{\rm e, foe}=10^{52}\ E_{\rm e, 51} f_{\rm b, -2}^{-1} t_{\rm e, min, 1}^{-1}$~\ergs.

Jets can deposit a significant energy into the stellar envelope on timescales of tens of seconds (i.e., $t \gtrsim t_{\rm e, min}$), but it is unlikely that they can produce the necessary mass of radioactive $^{56}$Ni to partially power the optical light curve of core-collapse SNe \citep[i.e., $\sim 0.02\,M_{\sun}-0.2\,M_{\sun}$,][]{hamuy_2003, mueller_2017}. Being a by-product of the explosive silicon burning process, $^{56}$Ni can be synthesized under specific temperature and density conditions \citep[e.g.][]{woosley_1973, woosley_1995, jerkstrand_2015}, which cannot be easily met once the ejecta has expanded far from the central engine.

Although failed jets cannot be the sole mechanism that drives the SN explosion, a picture emerges where they may be an important component of this process \citep[e.g.][]{lazzatietal2012, sobacchi2017, piranetal2017}. The limitation on jet-driven explosions stems from the energetic requirement of injecting $\sim 10^{51}$~erg and not from a lack of engines at these lower luminosities (see dash-dotted line in Fig.~\ref{fig:fig3}). 
The number of failed GRB engines with $L_{\rm e} < L_{\rm e,*}$ can be written as
$N_{\rm f}(\Le)\approx \left(\Le/L_{\rm e,*}\right)^{-\beta_L}$, where a normalization of one engine at luminosity $L_{\rm e,*}$ was assumed.  The true rate of failed engines capable of injecting $\sim 10^{51}$ erg at typical $\hat{z}=2$ can be then estimated as:
\eqb 
R_{\rm f}(\hat{z}) \!\! & \approx  & \!\! N_{\rm f}(L_{\rm e, foe}) R_{\rm GRB}(\hat{z}) f_{\rm b}^{-1}\\ 
\!\! &  \approx & \!\! 10^5 \, f_{\rm b,-2}^{0.4} \left(\frac{\eta_{-1} E_{\rm e, 51}}{t_{\rm e, min,1} L_{*, 52.5}}\right)^{-1.4}\!\!\!\!{\rm Gpc}^{-3} {\rm yr}^{-1} 
\eqe 
where $R_{\rm GRB}(\hat{z})\approx 10$~Gpc$^{-3}$~yr$^{-1}$ is  the observed (i.e., uncorrected for beaming) collapsar GRB rate \citep[][]{wanderman_piran2010, sobacchi2017}. We also used $\beta_L=1.4$ and assumed a redshift independent engine luminosity function. Interestingly, this order-of-magnitude prediction is consistent with inferred rates of Type Ib/c SNe at $\hat{z}\sim 2$ \citep[e.g.][]{dahlen2004, dahlen2012, cappellaro2015, strolger2015} suggesting an active role of the central engine in all Ib/c SNe \citep[see also][]{sobacchi2017, bear2017}. 

Our results, although consistent with those by \cite{sobacchi2017}, are derived with a very different method. The number 
and energetics of our failed jets, as well as the minimum engine timescale are naturally obtained by matching 
the distribution of successful engines with the GRB luminosity function, instead of relying on an extrapolation 
of the central engine duration distribution to an (almost) arbitrary value.

Our analysis was performed independently of the nature of the GRB central engine, which can be either a newly born black hole or a magnetar (i.e., a rapidly spinning highly magnetized neutron star), see, e.g., \cite{kumarandzhang2015} for a review on GRB central engines. Our predictions for the energetics of central engines are summarized in Fig.~\ref{fig:fig3} and can be used to constrain their nature. We showed that the beaming-corrected energy of a typical GRB engine that is active for $\temin \sim 10$~s is a few $\times 10^{52}$~erg, see also equation (\ref{eq:Ees-num}). This might be already in tension with the magnetar scenario, although recent studies cast doubt on previous estimates on the maximum energy extractable in the magnetar model \citep{brianetal2015}. In addition, the distributions of the engine duration and luminosity are expected to differ between models of the central engine. In the magnetar scenario, for example, where the luminosity and duration of the engine depend both on the magnetic field and spin period of the magnetar (e.g., \citealp{usov1992}), the distributions of $L_{\rm e}$ and $t_{\rm e}$ will be correlated. A careful comparison of our model predictions against several observational constraints, such as the GRB distribution  on the $L-T_{90}$ plane, will be the subject of a follow-up study.

\section{Conclusions}
The propagation of the jet through the envelope of a collapsing star, which acts as a filter of less powerful jets, may explain 
the observed GRB luminosity function and the GRB duration distribution. Our work helps to better understand the properties of long GRB central engines, while it provides interesting hints for the jet-driven SN scenario. 

\section*{Acknowledgements} 
We thank the anonymous referee for a constructive report. We also thank P. Beniamini, B. Metzger, and E. Nakar for useful comments on the manuscript. We thank Georgios Vasilopoulos for useful discussions. We acknowledge support from NASA through the grants NNX16AB32G and NNX17AG21G issued through the Astrophysics Theory Program. We also acknowledge support from the Research Corporation for Science Advancement’s Scialog program. 


\bibliographystyle{mnras} 
\bibliography{grb.bib} 
 
\end{document}